 \def\Title{Indistinguishability and the external correlation of mixtures}
 \def\arXiv{quant-ph/0308160v2}
 \def\Abstract{%
Experimental evidence, the heuristics of indistinguishability, and its logical
inconsistency with quantum formalism all argue against the existence of a quantum mixture
uncorrelated with the exterior, that is, argue for the postulate ``The state of a system
uncorrelated with its exterior is pure.'' This is shown to be equivalent with ``The state
of a system describable in terms of indistinguishable pure states is pure,'' and with ``The
state of the universe is pure''; further, it yields a quantitative expression of the
traditional relation of \wW information to partial coherence. It is concluded that all
mixtures are ``improper,'' the trace-reduction of a composite system's pure state.
}%
 \documentclass[aps,rmp,nobibnotes,nofootinbib,onecolumn,twoside,noshowpacs,draft]{revtex4}
 \usepackage[tbtags,sumlimits]{amsmath}
 \usepackage{amssymb}
 \makeatletter
 \def\p@section{}
 \def\p@subsection{}
 \makeatother

 \usepackage{xspace}

 \newcommand{\ie}{i.e., }
 \newcommand{\eg}{e.g., }
 \newcommand{\cf}{cf.\xspace}

 \newcommand{\RefSec}[1]{Sec.~\ref{#1}}
 
 \newcommand{\RefEqn}[1]{Eq.~\eqref{#1}}

 \newcommand{\QED}{\hfill\ensuremath{\square}\smallskip\smallskip}
 \newcommand{\Proof}{\par\noindent\emph{Proof:~}}
 \newcommand{\IFF}{\emph{{i\/f\/f}}\xspace}
 \newcommand{\abs}[1]{\ensuremath{\left\vert#1\right\vert}}
 \newcommand{\Sum}[1]{\ensuremath{\sum_{#1}}}
 \newcommand{\KDelta}[2]{\ensuremath{\delta_{{#1}{#2}}}}
 \newcommand{\set}[1]{\ensuremath{{\left\{\,#1\,\right\}}}}%
 \newcommand{\setsuch}[2]{\ensuremath{\big\{\,#1\,\big|\,#2\,\big\}}}%

 \newcommand{\Orj}[2]{\ensuremath{{\textstyle\bigvee}_{\!#1}\,{#2}}}

 \newcommand{\One}{\ensuremath{\mathbf{\displaystyle{1}}}\xspace}

 \newcommand{\Sys}[1]{\ensuremath{\mathcal{#1}}\xspace}
 \newcommand{\HS}[1]{\ensuremath{\mathcal{H}{}^{\Sys{#1}}}\xspace}
 \newcommand{\Trace}[2][]{\ensuremath{{\rm Tr}%
                    _{\Sys{#1}}\left\{\,{#2}\right\}}}
 \newcommand{\ket}[1]{\ensuremath{\vert\,{#1}\,\rangle}}
 \newcommand{\bra}[1]{\ensuremath{\langle\,#1\,\vert}}
 \newcommand{\braket}[2]{\ensuremath{\langle#1\,\vert\,#2\rangle}}
 \newcommand{\proj}[1]{\ensuremath{\ket{{#1}}\bra{{#1}}}}

 \newcommand{\PSI}[2]{\Psi^{\Sys{#1}}_{\,#2}}%
 \newcommand{\ketPsi}[2]{\ket{\PSI{#1}{#2}}\xspace}
 \newcommand{\bRho}{\pmb{\rho}}
 \newcommand{\Rho}[2]{\ensuremath{\bRho^{\Sys{#1}}_{#2}}\xspace}

 \newcommand{\State}[2]{\ensuremath{\sigma{}^{\scriptscriptstyle\Sys{#1}}_{#2}}\xspace}

 \renewcommand{\Pr}[2][]{\ensuremath{{\rm Pr}_{#1}\bigl(\,{#2}\,\bigr)}}
 \newcommand{\Prob}[3][]{\ensuremath{{\rm Pr}_{#1}\bigl(\,{#2}\bigm|#3\,\bigr)}}

 \newcommand{\Schrodinger}{Schr\"odinger\xspace}

 \newcommand{\wW}{\emph{welcher Weg}\xspace}
 \newcommand{\completeP}{complete$_\text{p}$\xspace}
 \newcommand{\completeV}{complete$_\text{v}$\xspace}

 \newcommand{\RID}{\emph{\bf ID}\xspace}
 \newcommand{\RIP}{\emph{\bf IP}\xspace}
 \newcommand{\RDD}{\emph{\bf DD}\xspace}
 \newcommand{\RDP}{\emph{\bf DP}\xspace}
 \newcommand{\PHP}{\emph{\bf HP}\xspace}

 \renewcommand{\b}[1]{\ensuremath{b_{#1}}\xspace}
 \newcommand{\p}[1]{\ensuremath{p_{#1}}\xspace}

 \newcommand{\SysS}{\Sys{S}}
 \newcommand{\SysM}{\Sys{M}}
 \newcommand{\SysSM}{\Sys{S\oplus M}}
 \newcommand{\SysE}{\Sys{E}}
 \newcommand{\SysSE}{\Sys{S\oplus E}}
 \newcommand{\SysME}{\Sys{M\oplus E}}

 \newcommand{\StateS}[1][]{\State{S}{#1}}
 \newcommand{\StateM}[1][]{\State{M}{#1}}
 \newcommand{\StateSM}[1][]{\State{SM}{#1}}

 \newcommand{\ketPsiS}{\ketPsi{\Sys{S}}{}}
 \newcommand{\ketPsiSM}{\ketPsi{\Sys{SM}}{}}
 \newcommand{\ketPsiSE}{\ketPsi{\Sys{SE}}{}}

 \newcommand{\RhoS}[1][]{\Rho{S}{#1}}
 \newcommand{\RhoM}[1][]{\Rho{M}{#1}}
 \newcommand{\RhoSM}[1][]{\Rho{SM}{#1}}

 \newcommand{\RhoSa}{\ensuremath{\bRho^{\SysS}_{1}}\xspace}
 \newcommand{\RhoSb}{\ensuremath{\bRho^{\SysS}_{2}}\xspace}

 \newcommand{\HSS}{\HS{S}}
 \newcommand{\HSM}{\HS{M}}
 \newcommand{\HSE}{\HS{E}}

 \usepackage{amsthm}
 \newtheorem{theorem}{Theorem}
 \newtheorem{lemma}{Lemma}

 \theoremstyle{definition}
 \newtheorem{definition}{Definition}

 \newcommand{\RefDefinition}[2][]{Def#1.~\ref{#2}}
 \newcommand{\RefTheorem}[2][]{Thm#1.~\ref{#2}}
 \newcommand{\RefLemma}[2][]{Lemma#1~\ref{#2}}

\begin{document}
 \makeatletter
 \def\ps@titlepage{%
   \renewcommand{\@oddfoot}{}%
   \renewcommand{\@evenfoot}{}%
   \renewcommand{\@oddhead}{\hfill\arXiv}
   \renewcommand{\@evenhead}{}}
 \makeatother

\title[Kirkpatrick -- \Title]{\Title}
 \author{K.~A.~Kirkpatrick}
 \email[E-mail: ]{kirkpatrick@nmhu.edu}
 \affiliation{New Mexico Highlands University, Las Vegas, New Mexico 87701}
\begin{abstract}
 \Abstract
\end{abstract}
 \maketitle
 \makeatletter\markboth{\hfill\@shorttitle\hfill}{\hfill\@shorttitle\hfill}\makeatother
 \pagestyle{myheadings}

\section{Introduction}%
``The concept of interfering alternatives is fundamental to all of quantum
mechanics''\citep{FeynmanHibbs65}. The relation of their interference to the
indistinguishability of these alternatives is equally fundamental: ``the loss of coherence
[interference] in measurements on quantum states can always be traced to correlations
between\dots the measuring apparatus and the system'' \citep{ScullyES89}. These
distinguishing correlations are traditionally called \wW (``which path'') information; the
lack of \wW correlations implies indistinguishability and coherence. Coherence, in quantum
mechanics, appears as the purity of the state descriptor, and incoherence as a mixed state
descriptor.%
\footnote{%
A quantal state is \emph{pure} \IFF the state operator is a projector.
} %

In classical probability, a preparation state is \emph{pure} if it is sharp
(dispersionless) in at least one variable.%
\footnote{%
As shown by example \citep{Kirkpatrick:Quantal}, even in classical probability the pure
state needn't be sharp in all variables.
} %
The mixed state, or \emph{mixture}, is a convex combination of pure states, and has
dispersion in all variables.

The mixture was introduced \emph{ad hoc} into quantum mechanics in direct analogy with the
classical mixture, as the mixing of pure state preparations ---
John~\citet{vonNeumann55tr}: ``if we do not even know what state is actually present ---
for example, when several states $\phi_1,\,\phi_2,\dots$ with the respective probabilities
$w_1,\,w_2,\dots$ constitute the description\dots,'' we have a mixture, represented by the
statistical operator $\Rho{}{}=\Sum{s}\,w_s\,\proj{\phi_s}$; Bernard~\citet{dEspagnat95}:
``An ensemble obtained by combining all the elements of several [pure state] subensembles
is a \emph{mixture}\dots''; Asher~\citet{PeresBook93}: ``[A] procedure in which we prepare
various pure states $\mathbf{u}_{\alpha}$ with respective probabilities $p_{\alpha}$''
leads to a mixture. However, each of these statements ignores the issue of
distinguishability: in quantum mechanics, the indistinguishable mixing of pure states
results in a pure state, not a mixture.

But alongside this \emph{ad hoc} introduction, the mixture arose deductively out of the
quantum formalism: With the simple requirement that the statistics of a proposition not be
changed by its conjunction with the trivial proposition in another system,
\citet{vonNeumann55tr} proved that \RhoS, the statistical descriptor of a subsystem \SysS
of a joint system \SysSM in the state \RhoSM, is uniquely given by the partial trace $
\RhoS=\Trace[M]{\RhoSM}$, which is a mixture if there is a distinguishing correlation
between the variables of \SysS and \SysM.

It is the purpose of the present paper to establish the converse ($\Trace[M]{\RhoSM}$ is
not a mixture if there is no such distinguishing correlation), thereby rejecting any
distinction between the mixture introduced \emph{ad hoc} and the mixture representing a
subsystem of a joint system: If a system's state is a mixture, that system is necessarily
correlated with another, external, system. \emph{There are no uncorrelated mixtures in
quantum mechanics}; mixtures that are assigned \emph{ad hoc} to the mixing of preparations
are in reality mixtures which arise from correlation with the exterior --- absent such
correlation the \emph{ad hoc} assignment is incorrect, and the state resulting from the
mixing is pure.

\RefSec{S:foundations} presents the necessary fundamentals of probability and correlation
(both in the classical and the quantal settings); it also presents (perhaps for the first
time) probabilistic definitions of indistinguishability independent of quantum mechanics.
\RefSec{S:Evidence} presents a micro-review of the experimental basis for the relation
between indistinguishability and coherence as stated by the well-known distinguishability
heuristics (stated here as \RIP and \RDP for combining preparation states
indistinguishably or distinguishably, respectively). \RefSec{S:anomaly} establishes that a
mixture uncorrelated with its exterior is an anomaly in quantum mechanics, which strongly
suggests the non-existence of uncorrelated mixtures: \emph{A system which is uncorrelated
with its exterior is in a pure state}; we denote this statement \PHP. (This is the
converse of the well known fact that \emph{a system in a pure state is uncorrelated with
its exterior}; \cf \RefTheorem{T:pureIsHermetic}.)

In \RefSec{S:Formalization}, we show  that \PHP is equivalent with a heuristic regarding
the indistinguishability of the states describing (supporting the state operator of) a
system. The implications of \PHP for the case of intermediate distinguishability are
discussed in \RefSec{S:General}. And finally, in \RefSec{S:Ignorance} we distinguish true
mixtures from the mixture-like mathematical expression used to estimate the state
resulting from an uncertain preparation.

\section{Fundamentals}\label{S:foundations}
Throughout this work, probability is understood statistically: that an event has a given
probability implies well-known statistical statements regarding the frequency of
occurrence of that event.

When dealing with a probabilistic physical system, it is necessary to distinguish the
\emph{probability state} (p-state) and the \emph{value state} (v-state): The p-state
carries the probabilities of all possible events (it is the card-count in a deck of cards,
the density matrix of a quantum mechanical system). The events themselves are described by
the occurrent values of the variables of the systems; this description is given by the
v-state.

We will consider physical systems which have several discrete-valued variables. The
probability that a variable $P$ of a system \SysS takes on the value \p{j} is denoted by
\Pr[\StateS]{\p{j}}, in which \StateS is the \emph{probability state} (p-state) determined
by the preparation of \SysS. (The value proposition $P=p_j$ will generally be denoted
simply by the value $p_j$.) A set of value propositions \set{\p{j}} is \emph{disjoint}
\IFF $\Pr[\StateS]{\p{j}\wedge\p{j'}}=\KDelta{j}{j'}$ for all \StateS (\ie all preparations),
and is \emph{\completeP{}}%
\footnote{We distinguish completeness in the probability sense, \emph{\completeP{\rm,}}
from completeness in the vector-space sense, \emph{\completeV{\rm.}}} %
\IFF $\Pr[\StateS]{\Orj{s}{\p{s}}}\!=\!1$ for all \StateS. If the set is disjoint,
\completeP-ness may be written $\Sum{s}\Pr[\StateS]{\p{s}}=1$. The value propositions of
a single variable are \completeP and disjoint. We may also consider sets of value
propositions which are not necessarily disjoint (that is, they are not the values of a
single variable); we will generally denote such by Greek rather than Roman letters:
\set{\phi_j}.

A system \SysSM may be considered the \emph{composite} of the two systems \SysS and
\SysM \IFF every p-state \StateSM of \SysSM implies unique p-states \StateS and \StateM
of \SysS and \SysM, respectively, such that
$\Pr[\StateS]{q}=\Pr[\StateSM]{q\wedge\text{T}}$ for every value proposition $q$ of \SysS
(with $\text{T}$ the trivial proposition of \SysM) --- and similarly with \SysS and \SysM
reversed. The probability of the occurrence of a value $p$ in system \SysS given the
occurrence of the value $q$ in system \SysM is the \emph{conditional probability}
$\Prob[\StateSM]{p}{a}={\Pr[\StateSM]{p\wedge a}}/\,{\Pr[\StateM]{a}}$.

In quantum mechanics a system \SysS is described in a Hilbert space \HSS. A value
proposition $q$ of the system \SysS is represented by the normalized vector
$\ket{q}\in\HSS$; vectors representing disjoint value propositions are orthogonal. A
p-state \StateS corresponds to a Hermitian statistical operator \RhoS with
$\textrm{Tr}\big\{\RhoS\big\}=1$, in terms of which
$\Pr[\StateS]{\phi}=\textrm{Tr}\big\{\RhoS\,\proj{\phi}\big\}$. (If \RhoS is a
projector, $\RhoS=\proj{\Psi^{\SysS}}$, and \ketPsiS also represents \StateS.) A
disjoint set \set{\p{j}} which is \completeP with respect to \StateS corresponds to an
orthogonal set \set{\ket{\p{j}}} which spans (is \completeV on) the support of \RhoS.
(\set{\ket{\p{j}}} may always be extended to an orthonormal basis of \HSS.) The joint
system \SysSM is described in the product space $\HSS\otimes\HSM$. The conjunction of
value propositions $\phi$ of \SysS and $\eta$ of \SysM is represented by the direct
product of their projectors: \proj{\phi\,\eta}. (As a notational shorthand, we denote
the direct product $\ket{\phi}\ket{\eta}$ as \ket{\phi\,\eta}.) Given a joint system's
statistical operator \RhoSM, the statistical operator of a subsystem \SysS is the
partial trace over the other system(s): $\RhoS=\textrm{Tr}_{\SysM}\big\{\RhoSM\big\}$.

For the ``entangled'' p-state $\sigma^{\Sys{AB}}$ represented by $\ket{\Psi^{\Sys{A}
\Sys{B}}}=\Sum{s}\psi_s\ket{\alpha_s\,\beta_s}$, we have the conditional probability
expression
\begin{equation}\label{E:Qconditional}
  \Prob[\sigma^{\Sys{AB}}]{a}{b}=%
     \big|\braket{a}{\Psi(b)}\big|^2,\quad\text{with }\ket{\Psi(b)}=%
     \frac{1}{N(b)}\Sum{s}\psi_s\,\braket{b}{\beta_s}\,\ket{\alpha_s}
\end{equation}
(where $N(b)>0$ normalizes \ket{\Psi(b)}). If the \set{\ket{\beta_j}} are orthonormal,
then
\begin{equation}\label{E:QcondON}
  \Prob[\sigma^{\Sys{AB}}]{a}{\beta_k}=\big|\braket{a}{\alpha_k}\big|^2.
\end{equation}

\subsection{Correlation and hermeticity}
The value propositions $\phi$ of \SysS and $\eta$ of \SysM are \emph{uncorrelated} in
the p-state \StateSM of \SysSM \IFF
$\Pr[\StateSM]{\phi\wedge\eta}=\Pr[\StateS]{\phi}\Pr[\StateM]{\eta}$. Systems which are
completely uncorrelated with their exteriors are an important part of quantum mechanics;
because there is no common term for this condition, I introduce the definitions
\begin{definition}\label{D:Hermetic}
A value of \SysS is \textbf{hermetic} \IFF it is uncorrelated with any value of any
system exterior to \SysS. A system is hermetic \IFF each of its values is hermetic.
\end{definition}
\begin{definition}
The \textbf{hermetic environment} of a system \SysS is the smallest hermetic system
which includes \SysS, less \SysS itself.
\end{definition}
\noindent(A system's hermetic environment always exists --- in the worst case, it would
consist of all other systems in the universe.)

\smallskip Several quantal results regarding hermeticity follow:
\begin{theorem}\label{T:HermeticFactors}
The quantal system \SysS is hermetic \IFF $\RhoSM=\RhoS\otimes\RhoM$ for every system
\SysM.
\end{theorem}
\Proof Sufficiency is obvious. Necessity: Hermeticity of \SysS implies
$\Pr[\StateSM]{\phi\wedge\eta}=\Pr[\StateS]{\phi}\,\Pr[\State{M}{}]{\eta}$ for every
value $\phi$ of \SysS and every value $\eta$ of every system \SysM; in quantum
mechanics, this is $\Trace[SM]{\RhoSM\,\proj{\phi}\otimes\proj{\eta}}=%
\Trace[S]{\RhoS\,\proj{\phi}}\Trace[M]{\RhoM\,\proj{\eta}}$ for every $\ket{\phi}\in\HSS$
and every $\ket{\eta}\in\HSM$;
$\Trace[S]{\mathbf{A}}\Trace[M]{\mathbf{B}}=\Trace[SM]{\mathbf{A}\otimes\mathbf{B}}$.
\QED

\begin{theorem}\label{T:pureIsHermetic}%
A quantal system whose probability state is pure is hermetic.%
\end{theorem}
\Proof The state operator of a composite system factors if one of the systems has a pure
p-state (\cf, \eg \textcite{BallentineBook98}, p.~219); apply
\RefTheorem{T:HermeticFactors}.\QED

\begin{theorem}\label{T:subHermetic}%
If the p-state of a composite quantal system is pure, and one of its subsystems has no
correlations with the rest of the composite system, then the p-state of that subsystem
is pure.
\end{theorem}
\Proof  Assume the state operator of the subsystem is not pure. Then it may be written as
a convex sum of orthogonal projectors, the eigenstates of some observable of that
subsystem. Because the composite system is pure, the \Schrodinger-HJW Theorem
(\citet{Schrodinger36}; see also \citet{Kirkpatrick:SchrHJW}) applies: those eigenstates
are correlated with some observable of the remainder of the composite system,
contradicting the assumption. \QED

\subsection{Indistinguishability}
That several values of a system \SysS are indistinguishable means that nothing in the world
external to \SysS reflects which value has occurred: the statistics of every value of the
exterior of \SysS must be independent of the various indistinguishable values of \SysS.
Thus,
\begin{definition}\label{D:Indistinguishability}%
The values \setsuch{\phi_{j}}{j\in\mathcal{I}}, a subset of the values of a system \SysS,
are \textbf{indistinguishable} \IFF $\Prob[\StateSM]{\eta}{\phi_{j}}$ is independent of
$j\in\mathcal{I}$, for every value $\eta$ of every system \SysM exterior to \SysS.
\end{definition}

\begin{theorem}\label{T:hermindist}%
All values of a hermetic system are indistinguishable.
\end{theorem}
\Proof Take \SysS to be any hermetic system; each value $\phi_j$ is hermetic, hence
uncorrelated with the exterior of \SysS. Thus, for each $j$,
$\Prob[\StateSM]{\eta}{\phi_j}=\Pr[\StateM]{\eta}$ for every $\eta$ of every exterior
\SysM: the \set{\phi_j} are indistinguishable.~\QED

\begin{theorem}
If a \completeP set of disjoint states are indistinguishable, they are hermetic.
\end{theorem}
\Proof \set{\p{j}} disjoint, \completeP: $\sum_t\Pr[\StateS]{\p{t}}=1$,
$\sum_t\Pr[\StateSM]{\p{t}\wedge\eta}=\Pr[\StateM]{\eta}$. If the \set{\p{j}} are
indistinguishable, then $\Prob[\StateSM]{\eta}{\p{j}}=f(\eta)$ for all $j$, hence
$\Pr[\StateSM]{\p{j}\wedge\eta}=\Pr{\p{j}}f(\eta)$; sum over $j$ to find
$f(\eta)=\Pr[\StateM]{\eta}$, hence \p{j} is uncorrelated with $\eta$. \QED

\subsection{Full distinguishability}
If several values of a system are fully distinguishable, there must be a dependable
external sign of which value has occurred; given the external sign, the statistics of
the system must be compatible with the corresponding occurrent value. Thus,
\begin{definition}\label{D:distinguishability}%
The values \setsuch{\phi_k}{k\in D} of a system \SysS are \textbf{fully distinguishable}
\IFF there exists, exterior to \SysS, a system \SysM which has a set of disjoint values
\setsuch{\b{k}}{k\in D} for which $\Prob[\StateSM]{q}{\b{k}}=\Pr[\phi_k]{q}$ for all
$k\in D$, for every value $q$ of \SysS (where \StateSM is the p-state of the joint system
\SysSM).
\end{definition}
\noindent(Only if the \set{\phi_j} are disjoint and \completeP may we define their full
distinguishability by the more obvious
``$\Prob[\StateSM]{\phi_j}{\b{k}}=\KDelta{j}{k}$.'')
\smallskip

The appendix contains several lemmas regarding distinguishability and
indistinguishability in quantum systems.

\section{Indistinguishability and preparation}\label{S:Evidence}%
Classically, the combining (``mixing'') of distinct pure-state preparations necessarily
results in a \emph{mixture}. In quantum mechanics, however, such mixing may result in a
pure p-state rather than a mixture, as we show in \RefSec{SS:coherentmix}.

It is often claimed that the mixing of independent preparations must necessarily result
in a mixture due to the indeterminacy of the phase of the several prepared pure p-states.
But this conflates a reduction of the visibility of coherence with an actual loss of
coherence, as we show in \RefSec{SS:invisible}.

In \RefSec{SS:IPDP} we present the distinguishability heuristics \RIP and \RDP.

\subsection{Evidence for the coherent mixing of preparations}\label{SS:coherentmix}%

A system is prepared randomly by one or another of several sources; each source prepares
the system in a distinct pure p-state. The intensities of the sources are sufficiently
low that never more than one system is available at a time. If the sources are aligned in
such a way that it is impossible in principle to determine from which source an
occurrence of the system arose --- hence impossible in principle to determine which pure
p-state the system was prepared in --- then this mixing of preparations yields a p-state
which is pure, a superposition of the several source preparation states, rather than a
mixture.

This particularly ``quantal'' behavior --- the indistinguishability, the lack of \wW
information, of alternative pure-state preparations leading to coherent superposition
--- has been recognized from the very earliest days of quantum mechanics. Over the years it
has been given quite direct experimental demonstration, particularly by Leonard Mandel's
group (below). The theoretical and experimental expressions of Marlan Scully's quantum
eraser (\citeauthor{ScullyDruhl82}, \citeyear{ScullyDruhl82}; \citeauthor{ScullyEW91},
\citeyear{ScullyEW91}; \citeauthor{KimEtAl99}, \citeyear{KimEtAl99}) fully exercise the
connection between the presence or absence of \wW information and the absence or presence
of coherent superposition.

\citet{PfleegorMandel67} demonstrated single-photon coherence with a pair of independent
equal-frequency lasers so arranged that, in their words, ``the localization of a photon at
the detector makes it intrinsically uncertain from which of the two sources it came.'' The
photon which has appeared at the detector is the result of mixing two pure-state
preparations (the emission of a photon from each laser), and the observed interference
shows that this mixing is at least partially coherent.

The 1991 experiment of Zou, Wang, and Mandel \citep{ZouWangMandel91,WangZouMandel91}
provides a clear demonstration of the influence of \wW information. Photons from a pair of
independent but phase-coherent sources (signal photons from a pair of coherently-pumped
downconversion crystals) travel on variable-differential-length paths to a detector; it is
extremely unlikely that photons from both sources are in the interferometer
simultaneously. The correlated idler photons leave the apparatus on a common path; a
variable-transmission filter causes the source of the idler photons, and hence of the
corresponding signal photons, to range from indistinguishable to fully distinguishable.
When indistinguishable, the signal photons are maximally coherent; as the
distinguishability increases, the coherence of the signal photons decreases, so that, for
full distinguishability, the signal photons are incoherently mixed. It is significant to
note that the idler photons are not ``detected'' in any way --- they, if not absorbed in
the transmission filter, pass out of the apparatus and travel through space until they
collide with whatever arbitrary matter might be on their path --- and that the signal
photons are not physically affected in any way by anything which happens to the idlers, in
particular not by changes in the transmissivity of the filter. Whether these independently
prepared signal photons combine incoherently or coherently depends only on the existence or
the nonexistence of their distant, distinct correlates.

\subsection{Invisible coherence}\label{SS:invisible}%
Consider two sources producing the system \SysS in the p-state $e^{i\theta_j}\ket{\psi_j}$,
with probabilities $w_j$ respectively; $j=1,\,2$. If these are indistinguishable
productions --- if there is no correlation of these two preparations with the exterior ---
then the p-state of the produced particle is necessarily
$\sqrt{w_1}\,e^{i\theta_1}\ket{\psi_1}+\sqrt{w_2}\,e^{i\theta_2}\ket{\psi_2}$, a pure
state, not a mixture. The coherence of this pure state may be seen in the interference in
the probability of passage through a \ket{x}-filter,
$w_1\,|\braket{\psi_1}{x}|^2 +w_2\,|\braket{\psi_2}{x}|^2%
+\sqrt{w_1w_2}\,\,|\braket{x}{\psi_2}\braket{\psi_1}{x}|\,\cos(\theta_2-\theta_1+\phi)$;
the cosine term is the interference. However, if the phase difference%
\footnote{%
Only the phase difference $\theta_2-\theta_1$ is physically meaningful.
} %
fluctuates widely, the time average of this interference term vanishes: This pure, but
time-dependent, p-state behaves as the mixture $w_1\,\proj{\psi_1}+w_2\,\proj{\psi_2}$
\emph{for all practical purposes} (FAPP). We see here the truth at the core of the claim
that the emerging ensemble is a mixture --- it is a ``mixture'' FAPP, though pure in
actuality.

Fluctuation of the phase difference is a matter of the experimental situation; if
experimental technique results in $\theta_2-\theta_1$ being constant during the
observation, interference will be visible and the coherence of the p-state will be
apparent. Pfleegor and Mandel repeatedly ``looked quickly,'' each look taken over such a
short time period that the phase difference was constant; Wang, Zou, and Mandel
time-stabilized the phase difference by coherent pumping. (A related discussion of this
matter is found in \citet{EnglertSW99b}.)

On the other hand, not only may a FAPP pseudo-mixture arise from time-fluctuating phases
in an indistinguishable mixing of preparations, but a true mixture may well arise out of
sloppy technique --- an interferometer on a wobbly table, say. In such case, the position
of the table is correlated with the phases of the photons, and directly averaging over the
phases, obtaining a mixture, is equivalent with tracing out the entangled environment (the
table). (See \citet{SternEtAl90} for an interesting treatment of this general question.)
In any discussion of fundamentals, we must take care to distinguish pure states which only
seem to be mixtures from true mixtures.

\subsection{Distinguishable and indistinguishable preparations}\label{SS:IPDP}%
If we prepare a system by randomly mixing several pure states, \emph{and do no more}, then
(in the interaction picture) the resulting p-state must be describable in terms of those
pure states:
\par\smallskip\noindent\textbf{Restriction on mixing-preparation.~}%
\emph{The p-state of a system prepared by randomly mixing a set of alternative pure
states \setsuch{\phi_j}{j\in\mathcal{D}} is supported by those pure states:
$\bRho=\sum_{tt'\in\mathcal{D}}w_{tt'}\,\ket{\phi_t}\bra{\phi_{t'}}$.}\smallskip
\par\noindent (That the support of the p-state is spanned by the several pure preparations
is implicitly assumed in all discussions of the fundamentals of quantum mixtures.)

Perhaps the clearest statement --- certainly the most consistent use --- of the principles
regarding \wW distinguishability and mixing preparations is found in
\citet{FeynmanVolIII}. The experiments reviewed in \RefSec{SS:coherentmix} strongly
support this statement, which may be expressed in terms of preparation states (rather than
Feynman's processes) as the two heuristics

\par\smallskip\noindent\textbf{Heuristic for Indistinguishable Preparations (\RIP).~}%
\emph{The p-state of a system prepared by randomly mixing a set of indistinguishable
alternative pure states \setsuch{\phi_j}{j\in\mathcal{D}}, each with probability $w_j$,
is pure,
$\bRho=\sum_{jj'\in\mathcal{D}}\psi_j\psi_{j'}^{\,*}\,\ket{\phi_j}\bra{\phi_{j'}}$
{\rm(}with $|\psi_j|^2=w_j${\rm)}.}\smallskip

\par\smallskip\noindent\textbf{Heuristic for Distinguishable Preparations (\RDP).~}%
\emph{The p-state of a system prepared by randomly mixing a set of fully distinguishable
alternative pure states \setsuch{\phi_j}{j\in\mathcal{D}}, each with probability $w_j$,
is a mixture, $\bRho=\sum_{j\in\mathcal{D}}w_j\,\proj{\phi_j}$.}\smallskip

Though the Heuristic for Indistinguishable Preparations must be considered a fundamental
part of quantum mechanics, it cannot be derived from the usual Hilbert-space
formalization of quantum mechanics. However, rather than postulating \RIP itself, we show
in the next section that \RIP strongly suggests another condition, \PHP, which itself
implies \RIP and is more suitable for statement as a postulate in a Hilbert-space
formalism.

\section{The hermetic mixture has no place in quantum mechanics}\label{S:anomaly}%
If the p-state of a hermetic system (\cf \RefDefinition{D:Hermetic}) is not pure, we
call it a \emph{hermetic mixture}. In this section we critically consider the place of
hermetic mixtures in quantum mechanics.

\subsection{The hermetic mixture cannot be created by mixing pure states}%
Suppose no hermetic mixtures already exist --- could we create one by mixing pure states?
The following theorem answers, No --- the resulting p-state would be pure or it would be
a non-hermetic mixture:
\begin{theorem}
Assuming \RIP and given the prior absence of hermetic mixtures in the environment, it is
impossible to construct a hermetic mixture by mixing pure-state preparations.
\end{theorem}
\Proof Let the preparation of the system \SysS vary randomly among a set of distinct
possible output p-states \set{\ket{\alpha_j}}. The corresponding p-states of the
environment must be pure: By hypothesis, there are no \emph{hermetic} mixtures, while if
the p-state of this ``environment'' were a \emph{correlated} mixture, then it must be only
a subsystem of the actual environment --- whose p-state therefore must be pure. Thus, when
\SysS is prepared in the p-state \ket{\alpha_j}, the p-state of its environment \SysE is a
pure state \ket{\eta_j}, so the p-state of \SysSE is \ket{\alpha_j\,\eta_j}. Now if these
p-states \set{\ket{\alpha_j\,\eta_j}} were not indistinguishable, there necessarily would
be a system \Sys{X} so that for each production the p-state of \Sys{S\oplus E\oplus X}
would be \ket{\alpha_j\,\eta_j\,\gamma_j}, the \set{\gamma_j} not collinear
(\RefLemma{L:relatedvectorsI}); but clearly then the environment of \SysS is truly
\Sys{E\oplus X}, contrary to assumption, so in fact the p-states
\set{\ket{\alpha_j\,\eta_j}} must be indistinguishable. Therefore, according to \RIP, the
p-state of \SysSE must be the pure state
$\ketPsiSE=\Sum{t}\gamma_t\,\ket{\alpha_t\,\eta_t}$ (for some set \set{\gamma_j}). If the
\set{\ket{\eta_j}} are all collinear, the state of \SysS is the pure state
$\ket{\Psi^{\SysS}}=\Sum{t}\gamma_t\,\ket{\alpha_{t}}$. If the \set{\ket{\eta_j}} are not
all collinear, then, utilizing the Schmidt decomposition of a pure state of a composite
system, we obtain $\ketPsiSE=\Sum{t}\psi_t\,\ket{p_t\,a_t}$, with both the \set{\ket{p_j}}
and the \set{\ket{a_j}} orthonormal, and with more than one $\psi_j$ non-vanishing. Then
the p-state of \SysS is $\RhoS=\Sum{t}\abs{\psi_t}^2\proj{p_t}$ --- a non-hermetic mixture.

This is exhaustive: the hermetic mixture cannot arise from any such construction. \QED

\subsection{Quantum mechanics provides no formalism regarding the mixing of hermetic mixtures}%
Suppose, if it were possible to prepare a system as a hermetic mixture, that we were to
mix several such preparations randomly --- for example, suppose \SysS were prepared in
the states \RhoSa and \RhoSb with the respective probabilities $w_1$ and $w_2$ --- what
would the resulting state be?

The intuitively obvious answer is, of course, $\RhoS=w_1\,\RhoSa+w_2\,\RhoSb$. The
equally obvious question arises immediately: How do we know this? And the only answer I
can find is: this is what we would expect of mixtures in classical probability ---
perhaps not the most convincing approach to take in quantal matters.

So let's look at this more carefully, using the specific example of an atomic-Young
double slit apparatus, with the p-states \ket{\p{1}} and \ket{\p{2}} representing passage
through each slit (again with probabilities $w_1$ and $w_2$).  Suppose that the p-state
of the environment is a mixture, \RhoM, uncorrelated with slit passage (so the slit
passages are indistinguishable); then the p-state of \SysSM is, for each slit passage,
the direct product of the pure state of \SysS with the mixed state of \SysM:
$\proj{\p{j}}\otimes\RhoM$. Following the ``intuitively obvious'' rule, the p-state of
\SysSM for the double-slit process is
$\RhoSM=w_1\,\proj{\p{1}}\otimes\RhoM+w_2\,\proj{\p{2}}\otimes\RhoM$; thus
$\RhoS=\Trace[M]{\RhoSM}=w_1\,\proj{\p{1}}+w_2\,\proj{\p{2}}$  ---  an incoherent mixture
lacking any double-slit interference! The intuitively obvious rule leads to
contradiction with standard quantum mechanics. And it's rather obvious that there's no
un-intuitive rule that's going to save the situation: there's not enough information in
the specification of the problem to obtain
$\ketPsi{S}{}=\sqrt{w_1}\,\ket{\p{1}}+\sqrt{w_2}\,\ket{\p{2}}$.

The situation is quite different if \RhoM is a \emph{correlated} mixture: Suppose the
system \SysME is in the pure state \ket{\PSI{ME}{j}} when \SysS passes through slit~$j$,
with $\Trace[E]{\proj{\PSI{ME}{j}}}=\RhoM$. Then (with $|\alpha_j|^2=w_j$),
$\ketPsi{SME}{}=
  \alpha_1\,\ket{\p{1}}\ket{\PSI{ME}{1}}+
  \alpha_2\,\ket{\p{2}}\ket{\PSI{ME}{2}}$, and so
\begin{gather}
\begin{split}
  \RhoSM=\big(&w_1\,\proj{\p{1}}+w_2\,\proj{\p{2}}\big)\otimes\RhoM\\
  &+ \big(\alpha_1\alpha_2^*\,\ket{\p{1}}\bra{\p{2}}\otimes%
    \Trace[E]{\ket{\PSI{ME}{1}}\bra{{\PSI{ME}{2}}}}\,
  +\text{h.~c.}\big),
  \end{split}
\end{gather}
the second term being what is missing after application of the ``intuitively obvious''
rule. The p-state of the system is thus
\begin{equation}
  \RhoS=w_1\,\proj{\p{1}}+w_2\,\proj{\p{2}}+
  \big(\alpha_1\alpha_2^*\,\braket{\PSI{ME}{2}}{\PSI{ME}{1}}\,
  \ket{\p{1}}\bra{\p{2}}+\text{c.~c.}\big);
\end{equation}
if the \ket{\PSI{ME}{j}} are orthogonal (so the slit passages are distinguishable),
\RhoS is an incoherent mixture, while if they are collinear (so the slit passages are
indistinguishable), \RhoS is the projector of the pure state
$\alpha_1\,\ket{\p{1}}+\alpha_2\,\ket{\p{2}}$.

\subsection{Thus spake Ockham: The hermetic mixture does not exist}
The hermetic mixture (a) cannot be distinguished phenomenologically from the correlated
mixture, other than by establishing a negative --- the absence, in the particular case,
of an ancillary correlate; (b) is treated only partially and inconsistently by the
standard formalism of quantum mechanics; (c) cannot be created by known quantal
processes; and (d) plays only a metaphysical role in the discussion of unobservable
aspects of ``subensemble'' membership. That is to say, neither reason nor evidence ---
only a kind of superstitious intuition --- supports the physical existence of the
hermetic mixture. Scientific conservatism, as expressed by Ockham, requires that the the
hermetic mixture be excluded by postulating that \emph{all physically existent mixtures
are correlated with the exterior}. This, our central result, may be stated more simply as
\par\smallskip \noindent\textbf{Postulate of Hermetic Purity (\PHP).~}\emph{The p-state of a
hermetic quantal system is pure.}\smallskip

The following sections will explore the implications of this postulate, but several
results follow immediately:
\begin{theorem}
$\RhoSM=\RhoS\otimes\RhoM$ for every system \SysM exterior to \SysS \IFF \RhoS is a
1-projector.
\end{theorem}
\Proof Necessity: \RefTheorem{T:HermeticFactors} and {\PHP}.\ \  Sufficiency:
\RefTheorem[s]{T:HermeticFactors} and \ref{T:pureIsHermetic}. \QED

The concept of ``quantum state of the universe'' is controversial, particularly from a
positivist viewpoint; it is interesting, though, that
\begin{theorem}\label{T:PHPPUP}%
The purity of the p-state of the universe is equivalent with the \emph{Postulate of
Hermetic Purity}.
\end{theorem}
\Proof Assume \PHP; the universe is necessarily hermetic, hence pure. Conversely, assume
the p-state of the universe to be pure; taking the composite system in
\RefTheorem{T:subHermetic} to be the universe, that theorem becomes \PHP. \QED

\section{Distinguishability heuristics in terms of state descriptions}\label{S:Formalization}%
A set of states which span the support of the statistical operator are sufficient to
describe the state; we formalize this as
\begin{definition}
The p-state of \SysS is \textbf{described by} the states
\setsuch{\ket{\phi_j}\in\HSS}{j\in\mathcal{D}} \IFF
$\RhoS=\sum_{tt'\in\mathcal{D}}w_{tt'}\,\ket{\phi_{t}}\bra{\phi_{t'}}$ (where
$w_{tt'}={w_{t't}}^*$ and
$\sum_{tt'\in\mathcal{D}}\,w_{tt'}\,\braket{\phi_{t'}}{\phi_t}=1$).
\end{definition}

The Restriction stated in \RefSec{SS:IPDP} requires that the mixing of pure states must
result in a p-state described by these pure states. Further, hermeticity of a system
implies indistinguishability of its values (\RefTheorem{T:hermindist}), and hermeticity of
a system implies purity of the p-state (\PHP). This suggests a strengthening of the
indistinguishability heuristics from preparations to state descriptors, and suggests that
a system described by a set of indistinguishable p-states must have a pure state --- that
is, this suggests the
\par\smallskip\noindent\textbf{Heuristic for Indistinguishable Descriptors (\RID).~}%
\emph{The p-state of a system described by (supported by) a set of indistinguishable
alternative pure states is pure.}\smallskip

\RID establishes that if \emph{any} set of state descriptors is indistinguishable, then
all sets are; in fact, as we show next, \RID is equivalent to \PHP. (Although \PHP is
proposed as a postulate, the following discussion treats it as merely a proposition
whose logical relation with \RID, also taken as a proposition, is to be explored.)

\begin{theorem}\label{T:Equivalent}%
The \emph{Postulate of Hermetic Purity} is equivalent to the \emph{Heuristic for
Indistinguishable Descriptors} \emph{(\PHP$\Leftrightarrow$ \RID)}.
\end{theorem}
\Proof\par\noindent$\RID\Rightarrow\PHP$: Assume \RID. Take \SysS to be any hermetic
system, its p-state described by a set of vectors \set{\ket{\phi_j}\in\HSS}; because
\SysS is hermetic, the \set{\phi_j} are indistinguishable (\RefTheorem{T:hermindist}),
\RID requires the \set{\ket{\phi_j}} be linearly combined to get the state vector of
\SysS, a pure state: \PHP.
\smallskip

\par\noindent$\PHP\Rightarrow\RID$: Assume \PHP.  Consider a system \SysS described by
a set of indistinguishable pure states \setsuch{\phi_{j}}{j\in\mathcal{I}}; these
correspond to the set \set{\ket{\phi_{j}}\in\HSS}. Let \SysE be the hermetic environment
of \SysS. By \PHP, the p-state of \SysSE, \ketPsiSE, is pure; it can be expanded in terms
of direct products of the \set{\ket{\phi_j}} and any orthonormal set
\set{\ket{b_j}\in\HSE}:
\begin{equation}
 \ketPsiSE=\Sum{s\in\mathcal{I}, t}\gamma_{st}\,\ket{\phi_s\,\b{t}}.
\end{equation}

\ketPsiSE may be rewritten in terms of a linearly independent subset of
\set{\ket{\phi_j}}: Let $\mathcal{L}\subset\mathcal{I}$ be the index set of a maximal
linearly independent subset of \set{\ket{\phi_j}}, so
$\ket{\phi_j}=\Sum{s\in\mathcal{L}}\alpha_{js}\,\ket{\phi_s}$; for $j\in\mathcal{L}$,
$\alpha_{jk}=\KDelta{j}{k}$. Then
\begin{equation}
 \ketPsiSE=\Sum{s\in\mathcal{L},t}\hat\gamma_{st}\,\ket{\phi_s\,\b{t}}, \text{where }%
   \hat\gamma_{st}=\Sum{p\in\mathcal{I}}\gamma_{pt}\alpha_{ps},\;s\!\in\!\mathcal{L}.
\end{equation}

This may be written in the correlated form
\begin{equation}
 \ketPsiSE=\Sum{s\in\mathcal{L}}\,\mu_s\,\ket{\phi_s\,\lambda_s},\;\text{with }%
 \mu_j\,\ket{\lambda_j}=\Sum{t}\hat\gamma_{jt}\,\ket{\b{t}}.
\end{equation}
But, by \RefLemma{L:relatedvectorsI}(b), the \set{\ket{\lambda_j}} must be collinear:
$\ket{\lambda_j}=e^{i\theta_j}\ket{\Lambda}$ for all $j$. Thus the p-state of \SysS is
pure, the linear combination of the indistinguishable p-states:
\begin{equation}
 \ketPsi{\SysS}{}=\Sum{s\in\mathcal{L}}e^{i\theta_s}\mu_s\,\ket{\phi_s}=%
              \Sum{s\in\mathcal{I}}\psi_s\,\ket{\phi_s}
\end{equation}
(the \set{\psi_j} not uniquely determined): \RID. \QED

In parallel with \RID, we state the
\par\smallskip\noindent\textbf{Heuristic for Distinguishable Descriptors (\RDD).~}%
\emph{The p-state of a system described by (supported by) a set of fully distinguishable
alternative pure states \setsuch{\phi_j}{j\in\mathcal{D}} is a mixture,
$\sum_{t\in\mathcal{D}}w_t\,\proj{\phi_t}$.}

\begin{theorem}\label{T:Implies}%
The \emph{Postulate of Hermetic Purity} implies the \emph{Heuristic for Distinguishable
Descriptors} \emph{(\PHP $\Rightarrow$ \RDD)}.
\end{theorem}
\Proof  Assume \PHP, and a system \SysS with a \completeV set of fully distinguishable
values \set{\ket{\phi_j}\in\HSS}.  The hermetic environment of \SysS is \SysE; by \PHP,
the p-state of \SysSE is pure, \ketPsiSE. Using the Schmidt decomposition,
$\ketPsiSE=\Sum{s}\mu_s\,\ket{p_s\,a_s}$, with the \set{\ket{p_j}\in\HSS} and the
\set{\ket{a_j}\in\HSE} orthonormal. Expanding
$\ket{p_s}=\Sum{t}\gamma_{st}\,\ket{\phi_t}$, we obtain
\begin{equation}\label{E:correlated}
\ketPsiSE=\Sum{t}\mu_t\,\ket{\phi_t\,\lambda_t},\text{ with }
\mu_t\,\ket{\lambda_t}=\Sum{s}\gamma_{st}\psi_s\,\ket{a_s}.
\end{equation}
By \RefLemma{L:distinguishable}, $\RhoS=\Sum{t}w_t\,\proj{\phi_t}$: \RDD. \QED
\par\noindent If the \set{\ket{\phi_j}} are not linearly independent,
\RefLemma{L:relatedvectorsD}(b) does not apply, and the \set{\ket{\phi_j}} may be fully
distinguishable even though the associated p-states of the exterior (the
\set{\ket{\lambda_j}}) are not orthogonal.

\begin{theorem}\label{T:IDIP}
The Heuristic for Indistinguishable Descriptors implies the Heuristic for
Indistinguishable Preparations; the Heuristic for Distinguishable Descriptors implies the
Heuristic for Distinguishable Preparations \textnormal{(}$\RID\Rightarrow\RIP;
\RDD\Rightarrow\RDP$\textnormal{)}.
\end{theorem}
\Proof The p-state of a system prepared by randomly mixing a set of indistinguishable
alternative pure states is supported by those indistinguishable states, hence by \RID is
pure; the p-state of a system prepared by randomly mixing a set of fully distinguishable
alternative pure states is supported by those distinguishable states, hence by \RDD is
mixed.~\QED

Thus, postulating \PHP recovers the traditional distinguishability heuristics \RIP and
\RDP.

\section{The general situation of partial distinguishability}\label{S:General}%
The categories \emph{fully distinguishable} and \emph{indistinguishable}, although
disjoint, are not complete: the intermediate case of \emph{partial distinguishability}
is not dealt with by the traditional \wW heuristics. Let us see what follows from \PHP:

A system \SysS is described in terms of the linearly independent
\set{\ket{\phi_j}\in\HSS}. Let \SysE be the hermetic environment of \SysS, so \SysSE is
hermetic; by \PHP, the p-state of \SysSE is pure, \ketPsiSE. We may write this in terms
of the \set{\ket{\phi_j}} and any orthonormal basis \set{\ket{b_j}\in\HSE}:
$\ketPsiSE=\Sum{st}\gamma_{st}\,\ket{\phi_s\,\b{t}}$; defining the \set{\ket{\lambda_j}}
by $\mu_s\,\ket{\lambda_s}=\Sum{t}\gamma_{st}\,\ket{b_t}$, we obtain the expression
\begin{equation}\label{E:general}
\ketPsiSE=\Sum{s}\mu_s\,\ket{\phi_s\,\lambda_s},\text{ and }%
 \RhoS=\Sum{ss'}\mu_s\mu_{s'}^{*}\,\braket{\lambda_{s'}}{\lambda_s}%
   \,\ket{\phi_s}\bra{\phi_{s'}}.
\end{equation}

Since, by \RefLemma{L:relatedvectorsI}, indistinguishability of the \set{\phi_j} requires
the collinearity of the \set{\ket{\lambda_j}} while, by \RefLemma{L:relatedvectorsD},
full distinguishability of the \set{\phi_j} requires the orthonormality of the
\set{\ket{\lambda_j}}, a set \set{\ket{\lambda_j}} which is neither orthonormal nor
collinear is necessary and sufficient to the intermediate case of partial
distinguishability of the \set{\phi_j}. Thus \RefEqn{E:general} yields the entire range
of possibilities for the \set{\phi_j} between indistinguishability and full
distinguishability; only the extremes are accounted for by the distinguishability
heuristics.

An empirical failure of \RefEqn{E:general} would not point unambiguously to an error in
the heuristics \RID and \RDD; in contrast, such failure would directly falsify \PHP.
Because of this greater degree of falsifiability, \PHP is scientifically stronger than
these heuristics.

A particularly interesting situation of partial distinguishability is that in which the
\set{\ket{\lambda_j}} are, pairwise, either orthonormal or collinear --- as for an
incomplete ideal measurement. In this case, the \set{\phi_j} will divide into a number
of subsets of indistinguishable values, the subsets fully distinguishable from one
another.  Take, for example, a three-slit atomic Young apparatus with an ideal passage
detector at slit~1, where the \set{\phi_j} represent the passage through the slits and
the \set{\lambda_j} represent the passage detector: The activation of the detector is
\ket{d_1} and its non-activation is \ket{d_0}, with $\braket{d_1}{d_0}=0$; the
assumption of ideality of the passage detector gives us $\ket{\lambda_1}=\ket{d_1}$  and
$\ket{\lambda_2}=\ket{\lambda_3}=\ket{d_0}$. Detecting the atoms at the screen yields
the reduced-visibility interference
\begin{equation}
 \RhoS=|\mu_1|^2\,\proj{\phi_1}+%
 \left(\,\mu_2\,\ket{\phi_2}+\mu_3\,\ket{\phi_3}\,\right)%
 \left(\,\mu_2^*\,\bra{\phi_2}+\mu_3^*\,\bra{\phi_3}\,\right)
\end{equation}
due to partial distinguishability. We see (using \RefEqn{E:QcondON}) that selecting out
the atoms at the screen which arrived in anti-coincidence with the passage detector
results in full-visibility interference between $\phi_2$ and~$\phi_3$ due to their
indistinguishability. (Imaginative extension of the \wW heuristics may yield this result;
such extensions, however, are \emph{ad hoc} and limited to the case at hand.)

\section{The ``Ignorance Mixture'' is not necessarily a mixture}\label{S:Ignorance}
The preparation determines the p-state, from which (taking into account the further
action of ``the whole experimental arrangement'') all probability predictions arise.
That preparation may be a random mixing of several pure-state sub-preparations
\set{\phi_j} each with probability $w_j$  (\eg a polarizer whose orientation varies a
little due to random rotational vibration of the optical bench). Let us specify the case
(call it A) of fully distinguishable mixing; the resulting p-state, as we've discussed
at length, is the mixture $\bRho=\sum_s\,w_s\,\proj{\phi_s}$.

There is another situation (call it B), easily conflated with (A): the preparation is one
of several possibilities, \set{\phi_j}; we don't know which one, but we can assign a
probability (subjective or objective, depending on circumstances) $w_j$ to the
possibilities  (\eg a rigidly mounted polarizer whose orientation is set with limited
accuracy). In this case, we must guess at the p-state --- and  our best initial estimate
of the preparation is the mixture-like expression
$\bRho_{\text{est}}=\sum_s\,w_s\,\proj{\phi_s}$.

Observations on any number of systems produced as per (A) will be statistically
consistent with the preparation $\bRho=\sum_s\,w_s\,\proj{\phi_s}$; in contrast, after
the observation of even a few systems produced as per (B), it is likely that the results
will be more consistent with a different expression
$\bRho'_{\text{est}}=\sum_s\,w'_s\,\proj{\phi_s}$, with the \set{w'_j} more ``peaked'';
with more observations, this would be expected to converge with confidence to a pure
state \proj{\phi_J} for some (presently unknown) $J$.

Situation (B) is a true situation of ignorance: we simply do not know which preparation
\set{\phi_j} was used, and, as we observe the systems, the estimate $\bRho_{\text{est}}$
``collapses'' to a better estimate $\bRho'_{\text{est}}$, reducing our ignorance.
Ignorance is quite irrelevant to situation (A); perhaps we know, at each occurrence, in
which $\phi_j$ the system was prepared, perhaps not --- nonetheless the statistics of
the occurrences continue to be described by $\bRho=\sum_s\,w_s\,\proj{\phi_s}$.

$\bRho$ is the statistical descriptor of a mixture which arises (in quantum mechanics)
from correlation with the exterior; $\bRho_{\text{est}}$ is merely an estimator of the
true (pure and hermetic) p-state. The use of a mixture-like estimator
$\bRho_{\text{est}}$ does not contradict the conclusion of this paper that all true
mixtures in quantum mechanics are accompanied by external correlation.

\section{Summary and conclusion}
The Heuristics of Indistinguishable and Distinguishable Preparations (\RIP and \RDP) have
been accepted as a fundamental part of quantum mechanics from its earliest days. They,
and the intermediate case of partially distinguishable preparations, are directly and
strongly supported by experiment. We have shown that \RIP implies that the hermetic
mixture cannot be prepared, in the absence of neighboring hermetic mixtures, by mixing
pure-state preparations; further, we have seen that logical anomalies would arise within
the quantum formalism if hermetic mixtures were to exist. These results point to the
non-existence of the hermetic mixture. This result, which we have called \PHP, is
stronger than the traditional distinguishability heuristics (\RIP and \RDP), which are
expressed in terms of mixed preparations; \PHP is in fact equivalent with the stronger
heuristics \RID and \RDD, which are expressed in terms of state descriptors (basis
vectors).

We conclude that the statement \emph{If \SysS is hermetic then \RhoS is pure} (\PHP) is
a necessary postulate of quantum mechanics, the correct formalization of the relation
between \wW information and coherence. It follows that any quantum mixture is the
trace-reduction of a composite system's pure state --- an ``improper''
\citep{dEspagnat95} mixture --- and that every expression of a mixture in the form of a
convex sum of projectors implies a correlation of those projector-states to an ancillary
variable in another system.

Finally, a comment of possible interest to an area of current research: when calculating
entanglement in a bipartite mixture, it is always physically legitimate to treat the
problem as tripartite and pure (the third system unknown, but physically existent).

\appendix
\renewcommand{\theequation}{\Alph{section}\arabic{equation}}%
\setcounter{equation}{0}
\renewcommand{\thetheorem}{\Alph{section}\arabic{theorem}}
\renewcommand{\thelemma}{\Alph{section}\arabic{lemma}}
\setcounter{theorem}{0} \setcounter{lemma}{0}

\section*{}\setcounter{section}{1}
The lemmas of this appendix depend on definitions \ref{D:Indistinguishability} and
\ref{D:distinguishability} (of indistinguishability and full distinguishability,
respectively).

In each, the p-state \StateSM of \SysSM is the pure state $\ketPsiSM$.

\begin{lemma}\label{L:relatedvectorsI}
$\ketPsiSM=\Sum{t}\,\mu_t\,\ket{\phi_t\,\lambda_t}$.
\smallskip
\par\noindent\emph{(a)}~If the \set{\ket{\lambda_j}} are collinear, the
\set{\phi_j} are indistinguishable.
\par\noindent\emph{(b)}~If the \set{\ket{\phi_j}} are  linearly independent and
indistinguishable, the \set{\ket{\lambda_j}} are collinear.
\end{lemma}
\Proof
\par\noindent(a)~Write the collinear \set{\ket{\lambda_j}} as
$\ket{\lambda_j}=e^{i\theta_j}\,\ket{\Lambda}$ for all $j$; then
$\ketPsiSM=\ketPsi{S}{}\otimes\ket{\Lambda}$, with
$\ketPsiS=\Sum{t}\mu_t\,e^{i\theta_t}\,\ket{\phi_t}$. Because the p-state of \SysS is
pure, it is hermetic (\cf \RefTheorem{T:pureIsHermetic}); by \RefTheorem{T:hermindist}
the p-states \set{\phi_j} are indistinguishable.

\par\smallskip\noindent(b)~Using \RefEqn{E:Qconditional}, the expression, in quantum
terms, of the indistinguishability of the \set{\phi_j} in the p-state \ketPsiSM becomes
\begin{equation}\label{E:IndistChi}
  \big|\braket{\eta}{\chi_j}\big|^2\;\text{is independent of $j$ for all \ket{\eta}, where }%
  \ket{\chi_j}\equiv\frac{1}{N_j}\Sum{t}\mu_t\,\braket{\phi_j}{\phi_t}\,\ket{\lambda_t}.
\end{equation}
Thus the \set{\ket{\chi_j}} is collinear: for all $j$,
$\ket{\chi_j}=e^{i\theta_j}\ket{X}$.

Define $\mathcal{Z}\equiv\setsuch{\ket{x}\in\HSM}{\braket{x}{X}=0}$. Then, from
\RefEqn{E:IndistChi},
\begin{equation}
  \bra{\phi_j}\,\Big(\Sum{t}\mu_t\,\braket{x}{\lambda_t}\,\ket{\phi_t}\Big)=0%
    ,\;\text{for all $j$, for all $\ket{x}\in\mathcal{Z}$}.
\end{equation}
The vector in parentheses lies in the subspace spanned by the \set{\ket{\phi_j}} and is
orthogonal to all of them; it must therefore vanish. Since the \set{\ket{\phi_j}} are
linearly independent, that vanishing requires $\braket{x}{\lambda_j}=0$ for all $j$, for
all $\ket{x}\in\mathcal{Z}$; thus each $\lambda_j$ is collinear with $X$. \QED

\begin{lemma}\label{L:distinguishable} $\ketPsiSM=\Sum{t}\,\mu_t\,\ket{\phi_t\,\lambda_t}$.
\smallskip\par\noindent If the \set{\phi_j} are fully distinguishable, then there must
exist an orthonormal set \set{\ket{b_j}\in\HSM} such that
$\ketPsiSM=\Sum{t}\,\psi_t\,\ket{\phi_t\,b_t}${\rm;} if, further, the \set{\ket{\phi_j}}
are linearly independent, then $\ket{\lambda_j}=\ket{b_j}$ and $\mu_j=\psi_j$.
\end{lemma}
\Proof The \set{\phi_j} are fully distinguishable, thus by definition there must exist a
\completeP disjoint set \set{b_j} of \SysM in terms of which
$\Prob[\StateSM]{q}{b_j}=\Pr[\phi_j]{q}$ for all values $q$ of \SysS. Using
\RefEqn{E:Qconditional} with the \completeV orthonormal set \set{\ket{b_j}}
corresponding to the \set{b_j}, we have the quantum expression
\begin{equation}
\Prob[\StateSM]{q}{b_j}=|\braket{q}{\Psi_j}|^2,\; \text{ with }\;
\ket{\Psi_j}=\frac{1}{N_j}\Sum{t}\mu_t\braket{b_j}{\lambda_t}\ket{\phi_t}.
\end{equation}
Of course, $\Pr[\phi_j]{q}=|\braket{q}{\phi_j}|^2$. Full distinguishability then requires
that, for all \ket{q}, $|\braket{q}{\Psi_j}|^2=|\braket{q}{\phi_j}|^2$; then
$\ket{\Psi_j}=e^{i\theta_j}\ket{\phi_j}$, and we have
$N_j\,e^{i\theta_j}\,\ket{\phi_j}=\Sum{t}\mu_t\braket{b_j}{\lambda_t}\ket{\phi_t}$.
Multiply this by $\otimes\ket{b_j}$ and sum on $j$. The \set{\ket{b_j}} are \completeV
and orthonormal, so $\Sum{j}\proj{b_j}=\One^{\SysM}$ and thus
$\Sum{j}\Sum{t}\mu_t\,\braket{b_j}{\lambda_t}\,\ket{\phi_t}\otimes\ket{b_j}=%
\Sum{t}\mu_t\,\ket{\phi_t}\otimes\ket{\lambda_t}$. Thus
\begin{equation}\label{E:sumlambdab}
 \Sum{j}N_j\,e^{i\theta_j}\,\ket{\phi_j\,b_j}=%
 \Sum{t}\mu_t\,\ket{\phi_t\,\lambda_t};
\end{equation}
set $\psi_j=N_j\,e^{i\theta_j}$. If the \set{\ket{\phi_{j}}}) are linearly independent,
$\psi_j\,\ket{b_j}=\mu_j\,\ket{\lambda_j}$.\QED

\begin{lemma}\label{L:relatedvectorsD}
$\ketPsiSM=\Sum{t}\,\mu_t\,\ket{\phi_t\,\lambda_t}$.
\smallskip
\par\noindent\emph{(a)}~If the \set{\ket{\lambda_j}} are orthonormal, the
\set{\phi_j} are fully distinguishable.
\par\noindent\emph{(b)}~If the linearly independent \set{\ket{\phi_j}} are fully
distinguishable, the \set{\ket{\lambda_j}} are orthonormal.
\end{lemma}
\Proof
\par\smallskip\noindent(a)~By \RefEqn{E:QcondON},
$\Prob[\Psi^{\Sys{SM}}]{q}{\lambda_j}=\Pr[\phi_j]{q}$.
\par\smallskip\noindent(b)~Apply \RefLemma{L:distinguishable} to the linearly independent
\set{\phi_j}.\QED

 \renewcommand{\refname}{\sc References}


\begin{thebibliography}{}

\bibitem[Ballentine, 1998]{BallentineBook98}
Ballentine, L.~E. (1998). {\em Quantum Mechanics -- {A} Modern Development}. World
Scientific, Singapore.

\bibitem[{d}'Espagnat, 1995]{dEspagnat95}
{d}'Espagnat, B. (1995). {\em Veiled Reality}. Addison-Wesley, Reading, MA.

\bibitem[Englert et~al., 1999]{EnglertSW99b}
Englert, B.-G., Scully, M.~O., and Walther, H. (1999). ``On mechanisms that enforce
complementarity,'' \eprint{quant-ph/9910037}.

\bibitem[Feynman and Hibbs, 1965]{FeynmanHibbs65}
Feynman, R.~P. and Hibbs, A.~R. (1965). {\em Quantum Mechanics and Path Integrals}.
McGraw-Hill, New York.

\bibitem[Feynman et~al., 1965]{FeynmanVolIII}
Feynman, R.~P., Leighton, R.~B., and Sands, M. (1965). {\em The {F}eynman Lectures on
Physics, {V}ol. {III}}. Addison-Wesley, Reading, MA.

\bibitem[Kim et~al., 2000]{KimEtAl99}
Kim, Y.-H., Yu, R., Kulik, S.~P., Shih, Y.~H., and Scully, M.~O. (2000). ``A delayed
choice quantum eraser,'' \emph{Phys. Rev. Lett.} {\bf 84}, 1--5,
  \eprint{quant-ph/9903047}.

\bibitem[Kirkpatrick, 2003a]{Kirkpatrick:Quantal}
Kirkpatrick, K.~A. (2003a). ``\,`{Q}uantal' behavior in classical probability,''
\emph{Found. Phys. Lett.}
  {\bf 16}(3), 199--224, \eprint{quant-ph/0106072}.

\bibitem[Kirkpatrick, 2003b]{Kirkpatrick:SchrHJW}
Kirkpatrick, K.~A. (2003b). ``The {S}chr{\"o}dinger-{HJW} {T}heorem,''
\eprint{quant-ph/0305068}.

\bibitem[Peres, 1993]{PeresBook93}
Peres, A. (1993). {\em Quantum Mechanics: {C}oncepts and Methods}. Kluwer, Dordrecht.

\bibitem[Pfleegor and Mandel, 1967]{PfleegorMandel67}
Pfleegor, R.~L. and Mandel, L. (1967). ``Interference of independent photon beams,''
\emph{Phys. Rev.} {\bf 159},
  1084--1088.

\bibitem[Schr{\"o}dinger, 1936]{Schrodinger36}
Schr{\"o}dinger, E. (1936). ``Probability relations between separated systems,''
\emph{Proc. Camb. Phil.
  Soc.} {\bf 32}, 446--452.

\bibitem[Scully and Druhl, 1982]{ScullyDruhl82}
Scully, M.~O. and Druhl, K. (1982). ``Quantum eraser: A proposed photon correlation
experiment concerning
  observation and `delayed-choice' in quantum mechanics,'' \emph{Phys. Rev. A}
  {\bf 25}, 2208--2213.

\bibitem[Scully et~al., 1989]{ScullyES89}
Scully, M.~O., Englert, B.-G., and Schwinger, J. (1989). ``Spin coherence and
{H}umpty-{D}umpty {III}. {T}he effects of observation,''
  \emph{Phys. Rev. A} {\bf 40}, 1775--1784.

\bibitem[Scully et~al., 1991]{ScullyEW91}
Scully, M.~O., Englert, B.-G., and Walther, H. (1991). ``Quantum optical tests of
complementarity,'' \emph{Nature} {\bf 351},
  111--116.

\bibitem[Stern et~al., 1990]{SternEtAl90}
Stern, A., Aharonov, Y., and Imry, Y. (1990). ``Phase uncertainty and loss of
interference: {A} general picture,''
  \emph{Phys. Rev. A} {\bf 41}, 3436--3448.

\bibitem[{v}on Neumann, 1955]{vonNeumann55tr}
{v}on Neumann, J. (1955). {\em Mathematical Foundations of Quantum Mechanics}. Princeton
University Press, Princeton, N. J. R.T. Beyer, tr. Originally published as
\emph{Mathematische Grundlagen der Quantenmechanik}, Springer, Berlin, 1932.

\bibitem[Wang et~al., 1991]{WangZouMandel91}
Wang, L.~J., Zou, X.~Y., and Mandel, L. (1991). ``Induced coherence without induced
emission,'' \emph{Phys. Rev. A} {\bf
  44}(7), 4614--4622.

\bibitem[Zou et~al., 1991]{ZouWangMandel91}
Zou, X.~Y., Wang, L.~J., and Mandel, L. (1991). ``Induced coherence and
indistinguishability in optical interference,''
  \emph{Phys. Rev. Lett.} {\bf 67}, 318--321.

\end{thebibliography}
\end{document}